\documentclass[fleqn,usenatbib]{mnras}

\usepackage[usenames,dvipsnames]{xcolor}
\usepackage{amsmath}
\usepackage{array}
\usepackage{xfrac}
\usepackage{multirow}

\usepackage[T1]{fontenc}
\usepackage{ae,aecompl}
\hypersetup{draft}

\title[Gas gradients on the mass--size plane]{SDSS-IV MaNGA: galaxy gas-phase metallicity gradients vary across the mass--size plane}

\author[Boardman et al.]{
N.~Boardman$^{1}$\thanks{E-mail: nick.boardman@astro.utah.edu},
G.~Zasowski$^{1}$,
J.~A.~Newman$^{2}$,
S.~F.~Sanchez$^{3}$,
A.~Schaefer$^{4},$\newauthor
J.~Lian$^{1}$,
D.~Bizyaev${^{5,6}}$,
N.~Drory${^7}$\\
$^{1}$Department of Physics \& Astronomy, University of Utah, Salt Lake City, UT, 84112, USA\\
$^{2}$Department of Physics \& Astronomy and PITT PACC, University of Pittsburgh, Pittsburgh, PA 15260, USA\\
$^{3}$Universidad Nacional Autónoma de México, Instituto de Astronomıa, A.P. 70-264, 04510, Mexico, D.F., Mexico\\
$^{4}$Department of Astronomy, University of Wisconsin-Madison, 475N. Charter St., Madison WI 53703, USA\\
$^{5}$Apache Point Observatory and New Mexico State University, P.O. Box 59, Sunspot, NM, 88349-0059, USA\\
$^{6}$Sternberg Astronomical Institute, Moscow State University, Moscow, Russia\\
$^{7}$McDonald Observatory, The University of Texas at Austin, 1 University Station, Austin, TX 78712, USA\\
}

\date{Accepted XXX. Received YYY; in original form ZZZ}
\pubyear{2019}
\begin{document} 
\label{firstpage}
\pagerange{\pageref{firstpage}--\pageref{lastpage}}
\maketitle

\begin{abstract}

Gas-phase abundances and abundance gradients provide much information on past stellar generations, and are powerful probes of how galaxies evolve. Gas abundance gradients in galaxies have been studied as functions of galaxies' mass and size individually, but have largely not been considered across the galaxy mass--size plane. Thus, we investigate gas-phase abundance gradients across this plane, using a sample of over 1000 galaxies selected from the MApping Nearby Galaxies at APO (MaNGA) spectroscopic survey. We find that gradients vary systematically such that above $10^{10}M_{\odot}$, smaller galaxies display flatter gradients than larger galaxies at a given stellar mass. This mass--size behaviour cannot be explained by instrumental effects, nor is it simply a reflection of known trends between gradients and morphology. We explore multiple possibilities for a physical origin for this pattern, though further work is needed to establish a firm physical interpretation.  

\end{abstract}

\section{Introduction}

Star formation within galaxies is fueled by the galaxies’ reservoirs of gas, which in turn contain imprints of the previous generations of stars. Thus, by studying the relative abundance of gas-phase elements, and by studying how they vary across galaxies, one can learn much about how galaxies form and evolve. 

Gas-phase metallicities in galaxies have been studied in great detail over the preceding decade, thanks in particular to large integral-field unit (IFU) spectroscopic datasets such as CALIFA \citep{sanchez2012}, SAMI \citep{croom2012}, and MaNGA \citep{bundy2015}. \citet{sanchez2012b} report a characteristic gas metallicity gradient of $\sim$0.1 dex per disc half-light radius. Subsequent work suggested a connection between these gradients and galaxies' morphologies \citep{sanchez2014,sm2016,sm2018}, as well as their masses \citep[e.g.][]{belfiore2017, carton2018}. Connections between gradients and galaxy sizes have also been reported \citep[though, see also][]{sm2018}. \citet{carton2018} report a connection between gas-phase metallicity gradients and galaxies' disc scale radii, such that larger galaxies display steeper size-scaled gradients on average size, while other studies have reported a steepening in physical gas metallicity gradients (units of dex/kpc) with decreasing galaxy size \citep{ho2015,bresolin2019}.

However, galaxy masses and sizes have long been known to be linked, such that more massive galaxies are on average more extended \citep{shen2003}. A galaxy's position along the mass--size relation is also connected to its morphology, such that later-type galaxies have larger sizes on average at a given stellar mass \citep{fl2013}. It is therefore more meaningful to consider galaxies' properties in terms of mass and size together. Such an approach been used before to assess IFU galaxy samples, but has largely focused on galaxies' stellar populations \citep[e.g.][]{mcdermid2015,scott2017,li2018}.

Here, we study the variation of gas-phase metallicity gradients and nitrogen abundance gradients along the mass--size plane, using measurements from the SDSS-IV MaNGA spectroscopic survey. We present the galaxy sample in \autoref{sample}, our results in \autoref{results}, and a discussion with conclusions in \autoref{discussion}.

\section{Sample and data}\label{sample}

We select a sample of galaxies from the MaNGA survey, which forms part of the Sloan Digital Sky Survey~IV \citep[SDSS-IV;][]{blanton2017}. Many of the measurements here are obtained from the MaNGA Pipe3D \citep{sanchez2016,sanchez2016b,sanchez2018} value added catalog (VAC)\footnote{available for galaxies in MPL-7 and below at \url{https://www.sdss.org/dr16/manga/manga-data/manga-pipe3d-value-added-catalog/}}, which contains absolute values and gradients of properties relating to galaxies' stellar and gaseous components; we use the MaNGA Product Launch 10 (MPL-10) version of this VAC. Measurements from MaNGA data are also available from the MaNGA Data Analysis Pipeline \citep[DAP;][]{belfiore2019,westfall2019}, and can be viewed and downloaded via the Marvin interface \citep{cherinka2019}.

The MaNGA observations were taken with the BOSS spectrographs \citep{smee2013} on the 2.5~m Sloan telescope at Apache Point Observatory \citep{gunn2006}. The MaNGA sample consists of roughly 10,000 galaxies selected to possess a wide range of morphologies and a roughly flat distribution in log-mass, with redshifts in the range of approximately 0.01 to 0.15 \citep{yan2016b,wake2017}. The individual MaNGA IFUs contain hexagonal fibre bundles of 19--127 optical fibres, each with diameter 2$^{\prime\prime}$; the IFU size is chosen for a given galaxy based on the galaxy's visible extent on the sky, with a three-point dithering pattern to fully sample the field of view \citep{drory2015,law2015}. Observations are reduced and drizzled with the MaNGA Data reduction pipeline \citep[DRP;][]{law2016,yan2016a}. The final reduced datacubes consist of $0.5^{\prime\prime} \times 0.5^{\prime\prime}$ spaxels, with a median point-spread function (PSF) full-width at half-maximum (FWHM) of approaximately 2.5$^{\prime\prime}$ \citep{law2015}. The reduced MaNGA spectra have a wavelength range of 3600--10000~\AA\ and a spectral resolution of $R \simeq 2000$ \citep{drory2015}. The first public MaNGA data release was included in SDSS Data Release 15 \citep[DR15;][]{aguado2019}, with further data (including DAP results) released in SDSS DR16 \citep{ahumada2020}.

We obtain from the Pipe3D VAC measurements of galaxies' gas-phase metallicity and nitogen abundance gradients in units of dex/$R_e$ (hereafter $\nabla$[O/H] and $\nabla$[N/O]). The gradients are measured between 0.5 $R_e$ and 2 $R_e$, with negative values indicating decreasing abundances with radius. $R_e$ is the elliptical Petrosian half-light radius, based on SDSS r-band photometry, from version 1.0.1 of the NASA-Sloan-ATLAS (NSA) catalog \citep{blanton2011}. The VAC includes [O/H] measurements from seven different calibrators, and includes [N/O] measurements from two different calibrators. In this paper, we focus on gas metallicities obtained from the \citet{marino2013} O3N2 calibrator, though we also tested values from the other six calibrators available on the VAC. For [N/O], we focus on the calibration of \citet{epm09}, though we also considered the alternative \citet{Dopita_2016_EmLineDiagnostic} calibrator.

\citet{belfiore2017} have previously demonstrated a secondary dependence on measured gradients with $R_e/PSF$, where the PSF is parametrised using the PSF FWHM. We likewise consider the $R_e/PSF$ of the galaxies. We also consider the morphology of a subset of galaxies by obtaining T-type values from the MaNGA Deep Learning Morphology VAC \citep[MDLM-VAC;][]{fischer2019}, which contains T-types from deep learning models for all MaNGA galaxies in MPL-7 and below.

We assemble a parent sample as follows. We first select from the Pipe3D VAC all galaxies that are part of the main MaNGA sample; this includes galaxies in the Primary and Color Enhanced samples (hereafter the ``Primary+'' sample) and the Secondary sample, which are observed out to approximately 1.5~$R_e$ and 2.5~$R_e$, respectively. In the case of duplicate galaxy observations, we use the observation with the highest combined signal-to-noise (S/N)\footnote{calculated as $(S/N)^2 = (S/N)_{\rm red}^2 + (S/N)_{\rm blue}^2$, where the latter variables describe the combined signal-to-noise in the red and blue cameras, respectively; see \citet{yan2016b} for more information.}. We restrict to galaxies with elliptical Petrosian axis ratios no lower than 0.6, to avoid edge-on galaxies. We then cross-match these galaxies with the GSWLC-2X catalog \citep{salim2018}, from which we obtain galaxy stellar masses derived from spectral energy distribution fits to combined ultraviolet-optical-infrared photometry; we rescale these masses to a \citet{kroupa2001} initial mass function (IMF). Restricting only to galaxies with listed stellar mass, we obtain a parent sample of 5458 galaxies.

From the parent sample, we assemble a final sample with which to study gas-phase abundance gradients. We select all galaxies with O3N2 gas-metallicity gradients listed in the Pipe3D VAC, obtaining 2548 in all. We then restrict to those galaxies with $\nabla$[N/O] errors of less than 0.1 dex/$R_e$ for both [N/O] calibrators; we found such a cut to be necessary due the significant amounts of scatter in the  $\nabla$[N/O] measurements.  This produces a final sample of 1679 galaxies, of which 824 have available T-type values. Our sample is overwhelmingly dominated by galaxies with significant ongoing star-formation, due to the requirement for observable gas-phase metallicity gradients. From the GSWLC-2X catalog, 1634 (97\%) of the galaxies have specific star-formation rates (sSFRs) greater than $10^{-11}$~yr$^{-1}$, with only one possessing a sSFR below $10^{-11.5}$~yr$^{-1}$.

\begin{figure}
\begin{center}
	\includegraphics[trim = 1cm 9cm 1cm 3cm,scale=0.42]{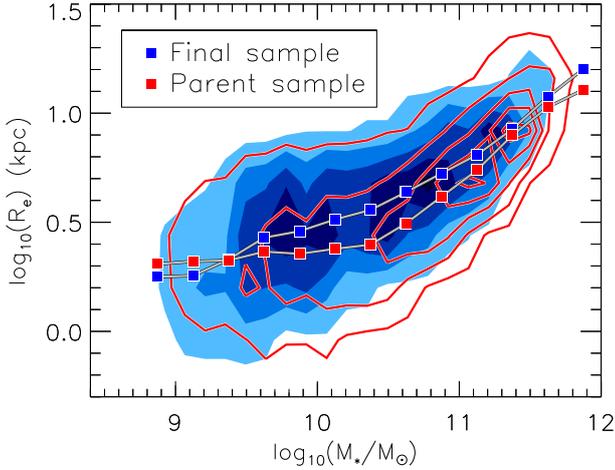}
	\caption{Mass--size relation of the final selected galaxy sample (blue) and the full parent sample (red). The contour boundaries mark 10\%, 30\%, 50\%, 70\% and 90\% of the highest-count bin of a given sample. The filled squares show the median size in logarithmic mass bins ($\Delta (\log{M_*})=0.25$), with the solid gray lines showing the interpolated mass-size relations.}
	\label{masssize_prelim}
	\end{center}
\end{figure}

In \autoref{masssize_prelim}, we present the distribution of the final selected sample in the mass--size plane. For comparison, we also show the mass--size distribution of the parent sample. We plot for both sets of galaxies the median galaxy size at a given stellar mass; we calculate the median size in a series of mass bins, before interpolating between the bin centres to estimate the median mass-size relation at a given galaxy mass. We see that the final galaxy sample tends towards larger average sizes than the parent sample at most masses. This is to be expected, due to our sample being dominated by star-forming galaxies.

\section{Gradients in the mass--size plane}\label{results}

\begin{figure}
\begin{center}
	\includegraphics[trim = 1cm 1.5cm 1cm 15cm,scale=0.65]{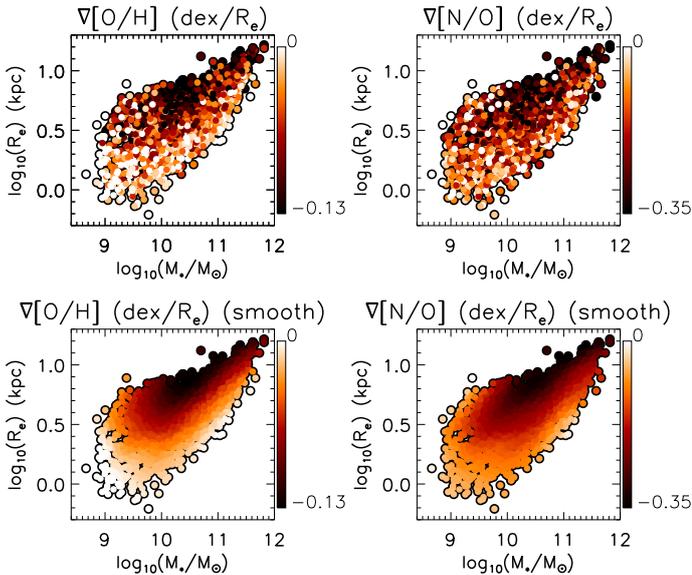}
	\caption{Mass--size relation of the MaNGA galaxy sample. Datapoints are colored by $\nabla$[O/H] (left) and $\nabla$[N/O] (right), both in dex/$R_e$, before (top row) and after (bottom row) LOESS smoothing.}
	\label{masssize_all_init}
	\end{center}
\end{figure}

In the top panels of \autoref{masssize_all_init}, we present the selected MaNGA sample in terms of $M_*$ and $R_e$, with datapoints coloured based on their $\nabla$[O/H] and $\nabla$[N/O] values. We note trends between the gas abundance gradients with stellar mass, such that the least massive galaxies display flatter gradients on average, similarly to what was previously reported in MaNGA galaxies by  \citet{belfiore2017} for gas metallicity gradients. We also note a trend with scale length, such that larger galaxies on average display steeper gradients, similarly to what \citet{carton2018} report for their sample of MUSE galaxies. 

To highlight the relationships along the mass--size plane, we apply the 2D Locally Weighted Regression \citep[LOESS;][]{cleveland1988} method as implemented in IDL\footnote{available from \url{http://www-astro.physics.ox.ac.uk/~mxc/software/}. We compute the smoothed value at each location using the closest 20\% of datapoints, with the \textit{rescale} keyword applied. We compute errors from the scatter in neighbouring points for the purpose of this calculation, rather than using the measurement errors from the Pipe3d VAC.} \citep{cappellari2013a}. This method has been used extensively in the past to uncover mean trends in integrated stellar population properties \citep{mcdermid2015,scott2017,li2018}, along with stellar population gradients in observations \citep{li2018} and simulations \citep{rosito2019}, but to our knowledge has not been employed to study gas-phase abundance gradients in observational samples. We show the results of LOESS smoothing in the bottom panels of \autoref{masssize_all_init}. Considering the residuals between smoothed and raw values of $\nabla$[O/H], we find a standard deviation of 0.04 dex/$R_e$; for $\nabla$[N/O], we find a standard deviation of 0.1 dex/$R_e$. Thus, the LOESS smoothing provides an accurate approximation of the underlying data. 
\section{Discussion}\label{discussion}

\subsection{Trends in massive galaxies}

We now consider more thoroughly the behaviour of galaxies with $M_* \geq 10^{10}M_\odot$, for which size appears an important predictor of measured gas-phase gradients; we hereafter refer to these 1019 galaxies as the ``massive subsample", with 480 having available T-type values. We explore whether the mass--size behaviour of these galaxies is physical or the result of an observational artifact, as well as assessing the impact of galaxies' morphologies. 

We define a new variable, $\Delta(R_e)$, that describes the logarithmic offset of a given galaxy's scale length from the median mass--size relation at its mass (as shown in \autoref{masssize_prelim}). We use here the median mass-size relation for the final selected sample. We have verified, however, that the results in this subsection are similar if we instead use the median mass--size relation from the parent sample; while $\Delta(R_e)$ tends towards higher values in this case, the {\it relative} differences in $\Delta(R_e)$ between galaxies --- which are key to our results --- change only slightly.

\begin{figure}
\begin{center}
	\includegraphics[trim = 1.cm 6.5cm 1cm 11cm,scale=0.7,clip]{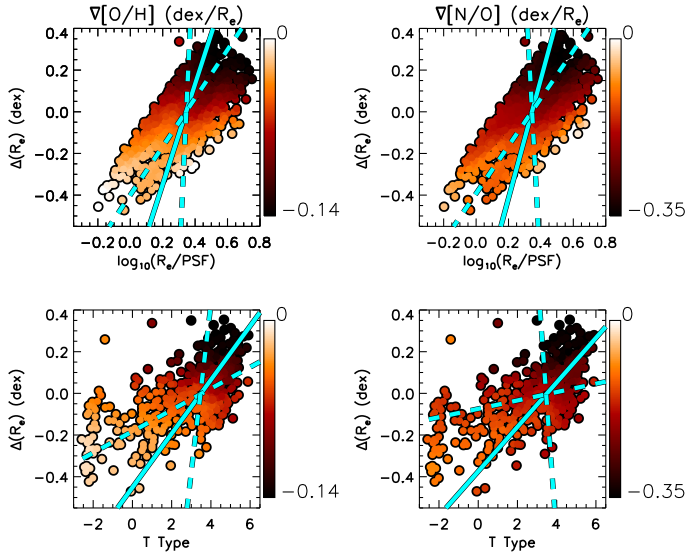}
	\caption{Maps of $\Delta R_e$ vs $R_e/PSF$ (top) and $\Delta R_e$ vs T-type (bottom), coloured by $\nabla$[O/H] (left) and $\nabla$[N/O] (right) after application of LOESS smoothing. The cyan lines show the best-fitting position angles, with the cyan dashed lines giving the $1\sigma$ errors.}
	\label{ohdisc}
	\end{center}
\end{figure}

In \autoref{ohdisc}, we plot the $\Delta(R_e)$ of the massive subsample as functions of $R_e/PSF$ and T-type, showing only galaxies included in MPL-7 in the latter case. We colour datapoints by $\nabla$[O/H] and $\nabla$[N/O], with LOESS smoothing applied. We find that $\Delta(R_e)$ is a more powerful predictor of $\nabla$[O/H] or $\nabla$[N/O]  than $R_e/PSF$, with gradients remaining related to $\Delta(R_e)$ at a given $R_e/PSF$. We also note a significant dependence on $\Delta(R_e)$ at a given  T-type for both $\nabla$[O/H] and $\nabla$[N/O]. As in \citet{fl2013}, we find later-type galaxies to have larger sizes on average at a given mass. Our results are consistent with previous reports of a connection between gas-phase metallicity gradients and morphology \citep{sm2016,sm2018}.

As an additional check, we parametrize the 2D relations with a ``position angle" that best encapsulates the trends in $\nabla$[O/H] or $\nabla$[N/O] along a given plane, shown with cyan lines in \autoref{ohdisc}. We obtain this angle with a procedure loosely adapted from the ``FIT\_KINEMATIC\_PA" IDL script \citep{kraj2006}. We calculate this angle from the y-axis of each 2D plane, after re-scaling the axes to have medians of 0 and standard deviations of 1; thus an angle close to 90$^\circ$ signifies a trend in the x-direction while an angle close to 0$^\circ$ signifies a trend in the y-direction. We perform this calculation by considering $\nabla$[O/H] or $\nabla$[N/O] as a function of a rotated axis, rotated from the y direction in steps of 1$^\circ$. We fit a 2nd order polynomial in each case, and define the position angle as that which minimises the $\chi^2$ between the data and the polynomial fit. 

We calculate angles for all three 2D parameter spaces shown thus far: the $M_*$--$R_e$ plane, the $\log R_e/PSF$--$\Delta (R_e)$ plane, and the T-type--$\Delta (R_e)$ plane. When considering $\nabla$[O/H], we obtain angles of $37^\circ \pm 4.5^\circ$, $20^\circ \pm 17^\circ$ and $31^\circ \pm 25.5^\circ$, respectively. For $\nabla$[N/O] we obtain angles of $35^\circ \pm 6^\circ$, $17^\circ \pm 20.5^\circ$ and $36^\circ \pm 39.5^\circ$, in turn. Thus, we again find $\Delta(R_e)$ to be a more fundamental predictor of measured gradients than $R_e/PSF$, with morphology also important. We quote errors as the 1$\sigma$ uncertainties after scaling errors on individual datapoints such that the reduced $\chi^2$ is equal to 1, with the error on the $\chi^2$ taken into account.

As a more quantitative analysis of different parameters' predictive power, we calculate the Spearman correlation coefficient $\rho$  between $\nabla$[O/H] and all other properties we have considered for the massive subsample, restricting to galaxies for which T-types are available. We find that $\Delta(R_e)$ yields the largest absolute $\rho$ ($-0.63$) when compared to $R_e/PSF$ ($-0.46$), T-type ($-0.53$), $R_e$ ($-0.39$) or $M_*$ ($0.09$); all but the last of these correlations have p-values of $p < 0.001$, with the final one having a p-value of 0.06. We see, however, that many parameters convincingly anti-correlate with $\nabla$[O/H] individually, emphasising the need to investigate parameters together. 

We next probe for secondary dependencies on $\nabla$[O/H] in the massive subsample. We fit a 2nd-order polynomial to $\nabla$[O/H] as a function of $\Delta(R_e)$ and calculate $\rho$ between the residuals and $R_e/PSF$, finding little correlation ($\rho = -0.09$, $p = 0.05$). If we fit a 2nd-order polynomial to $\nabla$[O/H] as a function of $\log (R_e/PSF)$, then the residuals correlate significantly with $\Delta(R_e)$ ($\rho = -0.38, p < 0.001$); this supports $\Delta(R_e)$ being a more powerful predictor of $\nabla$[O/H]. 

We obtain similar correlation coefficient results in terms of $\nabla$[N/O]. Considering $\nabla$[N/O] as a function of $\Delta(R_e)$, $R_e/PSF$ and T-type in turn, we find $\rho$ values of -0.47, -0.25 and -0.39, respectively. If we fit $\nabla$[N/O] as a function of $R_e/PSF$, then the residuals show a significant secondary dependency on $\Delta(R_e)$ ($\rho = -0.30$).

\begin{figure}
\begin{center}
	\includegraphics[trim = 1.cm 6.5cm 1cm 11cm,scale=0.7,clip]{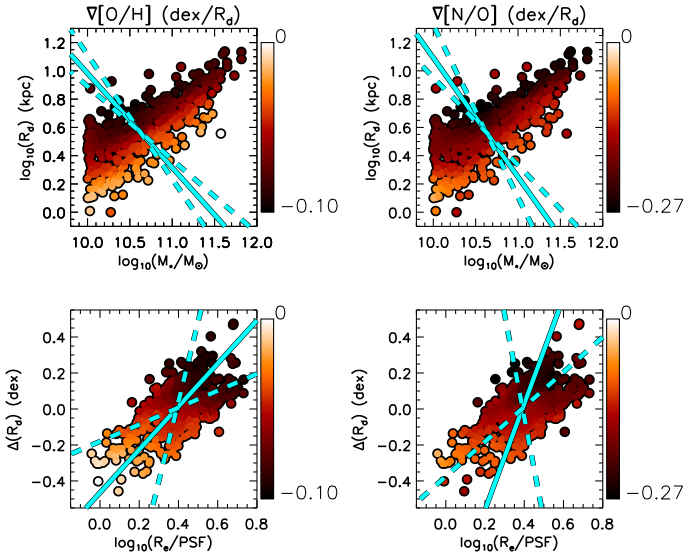}
	\caption{Maps of $M_*$ vs $R_d$ (top) and $\Delta R_d$ vs $R_e/PSF$ (bottom), coloured by $\nabla$[O/H] (left) and $\nabla$[N/O] (right) after application of LOESS smoothing and scaling of gradients into units of dex/$R_d$. The cyan lines show the best-fitting position angles, with the cyan dashed lines giving the $1\sigma$ errors.}
	\label{rddisc}
	\end{center}
\end{figure}

Finally, we consider the importance of our use of the half-light radius as a galaxy size parameter. We cross-match with the catalog of \citet{simard2011}, who provide bulge-to-total ratios (B/T) and disc scale lengths ($R_d$) for a large sample of SDSS galaxies. \citet{simard2011} calculate $R_d$ values from combined bulge+disc fits to $g$-band and $r$-band photometry, and calculate photometric B/T values for the two bands separately. We use the fits in which the bulge S\'ersic index treated as a free parameter, and we adopt the $r$-band B/T values. We restrict to galaxies in the massive subsample with B/T ratios of 0.3 or less. We parametrise galaxy size using $R_d$ and relative offsets from the mass--size plane using the parameter $\Delta (R_d)$ (analagous to $\Delta (R_e)$), and we scale abundance gradients into units of dex/$R_d$. We show the resulting LOESS-smoothed maps in \autoref{rddisc}. We find similar results as reported for the larger massive subsample, suggesting that our obtained mass--size behaviour does not depend on the adopted size measure.

 We thus find that the behaviour of the massive subsample along the mass--size plane is unlikely to be due to PSF effects alone. We further find that this behaviour is not driven purely by galaxy morphologies, and that the behaviour does not depend on the adopted size measure. These findings hold for both $\nabla$[O/H] and $\nabla$[N/O], and suggest the mass--size behaviour to be physical in nature. 
 
 We have performed additional checks with the massive subsample that we briefly summarise here. We have verified that our results are similar if we restrict to galaxies in the Primary+ or Secondary MaNGA samples. We have verified that our results are similar if we restrict our sample to galaxies observed with 91-fibre IFUs and larger, as an additional check against instrumental effects. We have also verified that our [O/H] results remain similar for all available Pipe3D calibrators. For [N/O], we note for the alternative \citet{Dopita_2016_EmLineDiagnostic} calibrator a stronger T-type dependence and a slightly weaker $\Delta(R_e)$ dependence ($\rho = 0.44$ and $\rho = 0.46$, respectively), though the qualitative mass--size behaviour remains unchanged.
 
We also note that the intrinsic mass--size trends are potentially slightly weaker than what observations imply, for two reasons. Firstly, $\Delta (R_e)$ is itself correlated quite strongly with $R_e/PSF$, meaning that any trend between measured gradients and $R_e/PSF$ would serve to strengthen the observed trend between measured gradients and $\Delta (R_e)$; we emphasise though that $\Delta (R_e)$ appears to be a strong predictor of the gradients \textit{at a given} $R_e/PSF$. Secondly, any errors in measured $R_e$ values will produce corresponding errors in the magnitudes of gradients, in the case when gradients are expressed as dex/$R_e$. However, typical $R_e$ errors are far too small to explain the trends that we observe. \citet{meert2013} find that half-light radii are recovered with scatters of approximately 5\%, for instance, from fits to mock SDSS galaxy images; this would imply an additional gradient error of around 5\%. Such an error is smaller than the typical reported measurement errors over much of the sample; for instance, if we consider all massive subsample galaxies with $\nabla$[O/H] of -0.02 and below, we find a median \% error of 11.4\% in $\nabla$[O/H] and 10.5\% in $\nabla$[N/O]. 

Our results here serve to further address an apparent tension in the literature, concerning the presence \citep[e.g.][]{sanchez2014, sm2018} or absence \citep[e.g.][]{belfiore2017,schaefer2020} of a characteristic gas metallicity gradient in nearby galaxies. It has been reported previously \cite[e.g.][]{sm2016} that gas metallicity gradients trend with morphology, such that earlier-type galaxies display flatter average gradients \citep[for reviews, see][]{sanchez2020a, sanchez2020b}. Here, we find that a galaxy's size acts as a predictor of measured gradients \textit{even at a given morphology}, suggesting both size and morphology to be relevant. To test this point, we advocate considering the gas metallicity gradients of other galaxy samples along the mass--size plane, along with considering sample galaxies' morphologies.

\subsection{Interpretation and conclusion}

To arrive at a physical understanding of the above results, we must ascertain how galaxy size connects to factors that have been previously reported to impact upon gas-phase abundances and gas-phase abundance gradients.

Gas-phase metallicity profiles and gradients were previously explored in MaNGA galaxies by \citet{belfiore2017}. Amongst other results, they report flat metallicity gradients for low mass ($M_* < 10^9M_\odot$) galaxies, consistent with what we find here; \citet{belfiore2017} argue this to show a need for strong feedback and gas-mixing, along with wind recycling (ejection and re-accretion of gas), in these galaxies. \citet{lian2019}, meanwhile, report little dependence on gradients with environment except for galaxies with masses below $10^{9.7}M_\odot$, at which point denser environments are associated with flatter gradients. 

For interpreting the mass--size behaviour of gradients in more massive galaxies, the density--metallicity relation is particularly relevant to consider. Larger galaxies are known to possess lower gas-phase metallicities than smaller galaxies at a given stellar mass \citep[e.g.,][]{ellison2008}, indicating a connection between metallicity and stellar mass density; in turn, galaxy populations display a local dependence between gas-phase metallicity and stellar mass surface density \citep[e.g.][]{ro2012,bb2016}. Given the connection between $\Delta (R_e)$ and stellar mass density, these dependencies should be explored further in the context of our results.

Another relevant factor is the fundamental metallicity relation \citep[FMR;][]{ellison2008,mannucci2010,ll2010}: at a given mass, galaxies with higher star-formation rates (SFRs) display lower integrated gas-phase metallicities on average. The majority of star-forming galaxies also display a local anti-correlation between SFR and gas metallicity once radial trends are removed \citep[e.g.][]{sm2019}, which is directly related to the FMR \citep{sa2019}. Since these local anti-correlations appear strongest in galaxies of stellar mass $M_* < 10^{10.5}M_\odot$ \citep{sm2019}, their importance in understanding mass--size trends amongst more massive galaxies is unclear. A local relation also exists between [O/H] and [N/O] abundances and star-formation efficiency \citep[SFE;][]{schaefer2020} and should likewise be considered further.

Looking to simulations, \citet{collacchioni2020} find gas-phase  metallicity gradients of EAGLE galaxies to anti-correlate with galaxies' gas accretion rates, with secondary anti-correlations at a given mass with neutral gas mass fraction and star-formation rate, as well as an overall secondary anti-correlation with stellar mass. \citet{hemler2020} similarly find an anticorrelation between gradients and mass for TNG50 $z \leq 2$ galaxies, but find gradients to become invariant with stellar mass if scaled by characteristic radii defined by galaxies' star-formation distributions. Merger histories have also been shown to impact upon gas metallicity gradients of simulated galaxies \citep{tissera2019}

To summarise, a number of factors are potentially relevant for interpreting our derived mass--size gradient trends, particularly with regards to the behaviour of the massive subsample. Local SFR-metallicity and density metallicity trends should be studied in this context, as should the importance of galaxies' SFE distributions. From simulations, gas accretion and merging should also be considered. 

We re-iterate at this point that our derived mass--size trends appear physical in nature: the behavior of the massive subsample cannot be ascribed purely to PSF effects, nor does it simply reflect differences in galaxies' morphologies. Further work is required to establish a detailed interpretation of the obtained trends. In addition, we encourage the behaviour of gas-phase metallicity along the mass--size plane to be explored in other IFU galaxy datasets.  

\section*{Data Availability}

All presented non-MaNGA data are publically available, as are MaNGA data for galaxies in MPL-7 and below from SDSS DR16. Roughly 50\% of the presented MaNGA galaxy sample is in MPL-7 or below. MaNGA galaxies in MPL-8 and later will be made available in SDSS DR17.  

\section*{Acknowledgements}

SFS is grateful for the support of a CONACYT grant CB-285080 and FC-2016-01-1916,
and funding from the PAPIIT-DGAPA-IN100519 (UNAM) project. Funding for the Sloan Digital Sky Survey IV has been provided by the Alfred P. Sloan Foundation, the U.S. Department of Energy Office of Science, and the Participating Institutions. SDSS-IV acknowledges support and resources from the Center for High-Performance Computing at the University of Utah. The SDSS web site is \url{www.sdss.org}.

SDSS-IV is managed by the Astrophysical Research Consortium for the Participating Institutions of the SDSS Collaboration including the Brazilian Participation Group, the Carnegie Institution for Science, Carnegie Mellon University, the Chilean Participation Group, the French Participation Group, Harvard-Smithsonian Center for Astrophysics, Instituto de Astrof\'isica de Canarias, The Johns Hopkins University, Kavli Institute for the Physics and Mathematics of the Universe (IPMU) / University of Tokyo, Lawrence Berkeley National Laboratory, Leibniz Institut f\"ur Astrophysik Potsdam (AIP),  Max-Planck-Institut f\"ur Astronomie (MPIA Heidelberg), Max-Planck-Institut f\"ur Astrophysik (MPA Garching), Max-Planck-Institut f\"ur Extraterrestrische Physik (MPE), National Astronomical Observatories of China, New Mexico State University, New York University, University of Notre Dame, Observat\'ario Nacional / MCTI, The Ohio State University, Pennsylvania State University, Shanghai Astronomical Observatory, United Kingdom Participation Group, Universidad Nacional Aut\'onoma de M\'exico, University of Arizona, University of Colorado Boulder, University of Oxford, University of Portsmouth, University of Utah, University of Virginia, University of Washington, University of Wisconsin, Vanderbilt University, and Yale University.

\bibliographystyle{mnras}
\bibliography{bibliography}

\label{lastpage}
\end{document}